\documentclass[reprint,prb,superscriptaddress,showpacs]{revtex4-2}%
\usepackage{graphicx}
\usepackage{xcolor}
\usepackage{amsmath}
\usepackage{lineno}
\usepackage{amsfonts}
\usepackage{amssymb}
\usepackage{framed}
\usepackage{float}
\usepackage{placeins}
\usepackage[normalem]{ulem}
\providecommand{\U}[1]{\protect\rule{.1in}{.1in}}
\makeatletter
\renewcommand*{\fnum@figure}{{\normalfont\bfseries \figurename~\thefigure}}
\renewcommand*{\@caption@fignum@sep}{\textbf{$|$ }}
\makeatother
\renewcommand{\figurename}{Fig.} 
\usepackage{hyperref}
\hypersetup{
    colorlinks=true,
    linkcolor=blue,
    filecolor=magenta,      
    urlcolor=cyan,
}
\begin{document}
\title{Tunable topological transitions in the frustrated magnet HoAgGe}


\author{Hari Bhandari}
\email{hbhandar@nd.edu}
\affiliation{Department of Physics and Astronomy, University of Notre Dame, Notre Dame, IN 46556, USA}
\affiliation{Stavropoulos Center for Complex Quantum Matter, University of Notre Dame, Notre Dame, IN 46556, USA}
\affiliation{Department of Physics and Astronomy, George Mason University, Fairfax, VA 22030, USA}
\author{Po-Hao Chang}
\affiliation{Department of Physics and Astronomy, George Mason University, Fairfax, VA 22030, USA}
\affiliation{Quantum Science and Engineering Center, George Mason University, Fairfax, VA 22030, USA}
\author{Resham Babu Regmi}
\affiliation{Department of Physics and Astronomy, University of Notre Dame, Notre Dame, IN 46556, USA}
\affiliation{Stavropoulos Center for Complex Quantum Matter, University of Notre Dame, Notre Dame, IN 46556, USA}
\author{Bence G. M\'arkus}
\affiliation{Department of Physics and Astronomy, University of Notre Dame, Notre Dame, IN 46556, USA}
\affiliation{Stavropoulos Center for Complex Quantum Matter, University of Notre Dame, Notre Dame, IN 46556, USA}
\author{L\'aszl\'o Forr\'o}
\affiliation{Department of Physics and Astronomy, University of Notre Dame, Notre Dame, IN 46556, USA}
\affiliation{Stavropoulos Center for Complex Quantum Matter, University of Notre Dame, Notre Dame, IN 46556, USA}
\author{John F. Mitchell}
\affiliation{Materials Science Division, Argonne National Laboratory, Lemont, IL 60439, USA}
\author{Igor I. Mazin}
\affiliation{Department of Physics and Astronomy, George Mason University, Fairfax, VA 22030, USA}
\affiliation{Quantum Science and Engineering Center, George Mason University, Fairfax, VA 22030, USA}
\author{Nirmal J. Ghimire}
\email{nghimire@nd.edu}
\affiliation{Department of Physics and Astronomy, University of Notre Dame, Notre Dame, IN 46556, USA}
\affiliation{Stavropoulos Center for Complex Quantum Matter, University of Notre Dame, Notre Dame, IN 46556, USA}

\begin{abstract}

The kagome lattice, known for its strong frustration in two dimensions, hosts a variety of exotic magnetic and electronic states. A variation of this geometry, where the triangular motifs are twisted to further reduce symmetry, has recently revealed even more complex physics. HoAgGe exemplifies such a structure, with magnetic and electronic properties believed to be driven by strong in-plane anisotropy of the Ho spins, effectively acting as a two-dimensional spin ice. In this study, using a combination of magnetization, Hall conductivity measurements, and density functional theory calculations, we demonstrate how various spin-ice states, stabilized by external magnetic fields, influence the Fermi surface topology. More interestingly, we observe sharp transitions in Hall conductivity without concurrent changes in magnetization when an external magnetic field is applied along a particular crystallographic direction, underscoring the role of strong magnetic frustration and providing a new platform for exploring the interplay between magnetic frustration, electronic topology, and crystalline symmetry. These results also highlight the limitations of a simple spin-ice model, suggesting that a more sophisticated framework is necessary to capture the subtle experimental nuances observed.
\end{abstract}
\maketitle

\section*{Introduction}\label{sec:1}

Magnetic frustration is widely known to give rise to a variety of exotic phases, such as quantum spin liquids and spin ice states, 
which host novel particles such as Majorana fermions and magnetic monopoles \cite{balents2010spin,castelnovo2008magnetic,motome2020hunting}. Among highly frustrated systems, kagome magnets stand out \cite{ghimire2020topology}. In a kagome lattice, corner-sharing equilateral triangles form perfect hexagons. When these triangles are rotated with respect to each other, they create what is called ``twisted kagome lattice'' \cite{Huang2023,zhao2024discrete}, reducing symmetry and increasing in-plane anisotropy. Additionally, in systems involving
rare earth elements, the crystal field splitting introduces an additional energy scale \cite{lee2023interplay}. In scenarios where the electronic bands near the Fermi energy are derived from the rare earth element, there is a unique opportunity to tune magnetism by manipulating anisotropies through external parameters such as magnetic fields. This tuning becomes even more compelling if the electronic bands exhibit topological features, such as flat bands and Dirac or Weyl points--key areas of interest in condensed matter physics.  


\begin{figure*}[ht]
\begin{center}
\includegraphics[width=.8\linewidth]{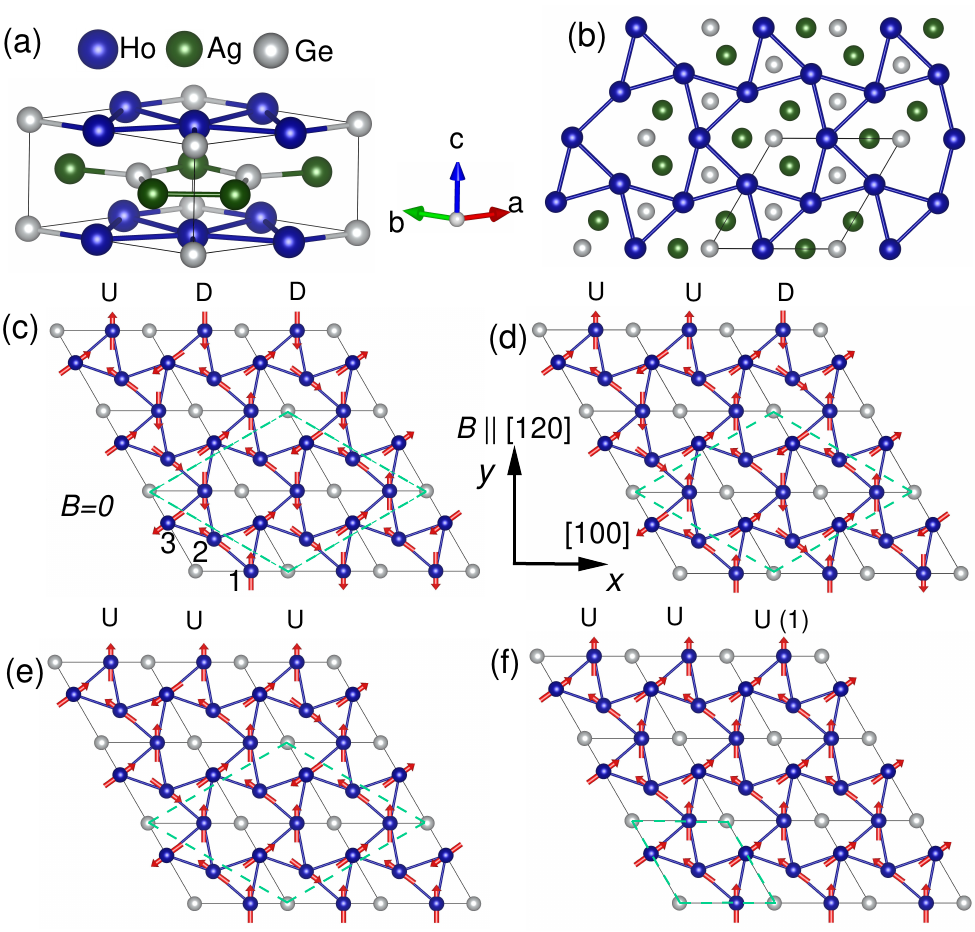}
    \caption{\small \textbf{Crystal and magnetic structures of HoAgGe.} \textbf{a}) A sketch of the crystal structure of HoAgGe. \textbf{b}) A view along the $c$-axis, highlighting the twisted kagome network of Ho atoms in the $ab$-plane. (\textbf{c} - \textbf{f}) The low-temperature magnetic structure of HoAgGe under different magnetic fields applied along the [120] crystallographic direction. The [120] direction, relative to the crystal plane, is indicated by the black arrow. Solid lines represent the crystallographic unit cells. Red arrows in panels \textbf{c}-\textbf{f} denote the spin directions. "U", and "D" refer to up, and down spins, respectively. Panel (\textbf{c}) shows the the ground state ($B=0$) UDD structure. The numbers 1, 2, and 3 correspond to Ho$_1$, Ho$_2$, and Ho$_3$ atoms, as defined in the main text. At a small magnetic field $B_1$, one down spin (D) of Ho$_1$ flips, leading to the UUD structure (panel \textbf{d}). As the magnetic field increases to $B_2$, another Ho$_1$ down spin flips, resulting in the UUU structure (panel \textbf{e}). At the saturated field, $B_3$, spins of all Ho$_3$ atoms flip, producing the UUU(1) structure (panel \textbf{f}). These magnetic structures were determined in a previous study \cite{zhao2020realization}. The green dashed lines represent the magnetic unit cell.}
    \label{Fig1}
    \end{center}
\end{figure*}

HoAgGe features a twisted kagome net formed by a rare earth element Ho \cite{GIBSON199634,morosan2005field}. It has a non-centrosymmetric crystal structure $P\bar{6}2m$, with alternating layers of Ho$_3$Ge and Ag$_3$Ge$_2$ along the [001] direction (Fig. \ref{Fig1}(a)). The equilateral triangles in the lattice are rotated by $\approx 15.6^{\circ}$ around the $c$-axis \cite{zhao2020realization}, forming the twisted kagome network shown in Fig. \ref{Fig1}(b). Previous studies have revealed that HoAgGe undergoes two successive antiferromagnetic transitions at 11 K ($T_{\text{N1}}$), and 7 K ($T_{\text{N2}}$) \cite{morosan2005field,zhao2020realization,roychowdhury2024enhancement,li2022low,deng2024local}. Below $T_{\text{N1}}$, the spins in Ho atoms partially order, while a fully ordered magnetic state emerges below $T_{\text{N2}}$ \cite{zhao2020realization}. 

Given the large single-site magnetic anisotropy energy (MAE) of the f-electrons, the low-temperature ground state has been interpreted as a $\sqrt{3}\times\sqrt{3}$ kagome spin ice, as shown in Fig. \ref{Fig1}(c) \cite{zhao2024discrete}.
There are three distinct types of Ho spins: Ho$_1$ with spins along [120], Ho$_2$ with a positive projection along [120], and Ho$_3$ with a negative projection along [120] as labeled in Fig. \ref{Fig1}(c). 

We refer to the ground-state phase, based on the orientation of the Ho$_1$ spins, as the up-down-down (UDD) phase. When an external field is applied along the [120] direction, the flipping of the two down-oriented Ho$_1$ spins results in the UUD, and UUU phases, as shown in Figs. \ref{Fig1}(d) and \ref{Fig1}(e). As the magnetic field increases further, Ho$_2$ spins flip, leading to the UUU(1) phase in Fig. \ref{Fig1}(f)]. These phases were identified as the 1-in-2-out or 2-in-1-out spin ice states in Ref. \cite{zhao2020realization}, where each spin flip causes a magnetization jump, creating 1/3 magnetization plateaus. 

Additionally, it has been shown \cite{zhao2024discrete} that within the first (1/3), and the second (2/3) plateaus, time-reversal-like meta-stable ice-rule states --- exhibiting identical magnetization but differing in magnetotransport properties --- are stabilized, presumably due to the twisting of the Ho-kagome-net. To the best of our knowledge, the magnetic phases developed under an external field along the [100] direction have not been previously studied, despite crystallographic differences between the two cases. Furthermore, the validity of the spin-ice model with infinite MAE has never been put to quantitative tests. There has also been no microscopic explanation proposed for the coupling of these metastable states or for the  appearance of the narrow (1/6) and (5/6) phases, identified in Ref. \cite{zhao2020realization}.

In this article, we investigate the impact of an in-plane magnetic field on the electronic band structure using magnetization, Hall effect measurements, and density functional theory (DFT) calculations. We find that the twisted kagome geometry leads to topologically nontrivial features such as quasi-2D bands, which have very narrow dispersion along a particular high-symmetry path in the Brillouin zone, and kagome Dirac crossings. Magnetic ordering into the ground state $\sqrt{3}\times\sqrt{3}$ supercell state results in folding of these bands, leading to a change in Fermi surface topology. External magnetic field-induced spin flipping that does not alter the supercell structure has little effect on the electronic structure, even with a substantial change in magnetization. However, when a spin flip transforms the supercell into the $1\times1$ structure, a significant change in Fermi surface topology occurs, causing an abrupt reversal in the Hall conductivity sign thereby regaining the quasi-2D and Dirac bands closer to the Fermi energy and with a significant spin polarization, a much-anticipated outcome of band structure manipulation by an external magnetic field. Even more intriguing, when the external magnetic field is aligned along the crystallographic $a$-axis, the Hall conductivity exhibits significant changes in Fermi surface topology without a corresponding change in magnetization. This anomalous behavior is attributed to magnetic field-induced frustration caused by the distortion of the perfect kagome geometry. These findings highlight the twisted kagome lattice as an important platform for exploring the interplay between magnetism and band topology.

\section*{Results}

\begin{figure}[ht]
\begin{center}
\includegraphics[width=.7\linewidth]{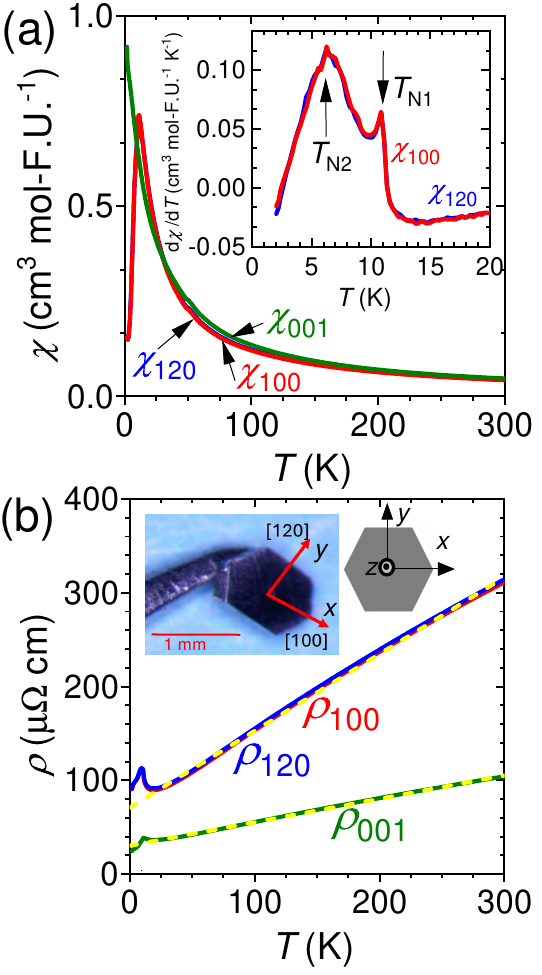}
    \caption{\small \textbf{Physical properties characterization of HoAgGe.} \textbf{a}) Magnetic susceptibility measured with an external magnetic field of 0.1 T along [100], [120], and [001] directions using the field-cooled protocol. \textbf{b}) Electrical resistivity as a function of temperature, $\rho(T)$, measured with current applied along [100], [120], and [001] directions. The yellow lines represent linear fits to $\rho(T)$ above 30 K.  The inset on the left shows an optical image of a polished crystal, illustrating crystallographic directions within the $ab$-plane. The inset on the right shows a sketch of the hexagonal geometry defining the $x$, $y$, and $z$ axes.}
    \label{Fig2}
    \end{center}
\end{figure}

\subsection*{Magnetic susceptibility and resistivity}

The magnetic susceptibility ($\chi$) and longitudinal resistivity ($\rho$) along the three crystallographic directions [100], [120] and [001] are shown in Fig. \ref{Fig2}. A clear antiferromagnetic transition at 11 K ($T_{\text{N1}}$) is observed in $\chi_{100}$ and $\chi_{120}$, while $\chi_{001}$ shows a steady increase down to 1.8 K, the lowest temperature measured [Fig. \ref{Fig2}(a)]. Nevertheless, $T_{\text{N1}}$  is detected in the resistivity along all three directions [Fig. \ref{Fig2}(b)]. Furthermore, both $T_{\text{N1}}$ and $T_{\text{N2}}$ are evident in the derivative of susceptibility [see the inset of Fig. \ref{Fig2}(a)] and  of resistivity (Supplementary Figure 1). This is consistent with previous studies \cite{morosan2005field,zhao2020realization,roychowdhury2024enhancement,li2022low,deng2024local}. The in-plane resistivities both along [100] ($\rho_{100}$) and [120] ($\rho_{120}$) are almost identical, and in the entire temperature range larger than the out-of-plane resistivity $\rho_{001}$, suggesting an anisotropic Fermi surface \cite{bhandari2304magnetism}. Additionally, in the non-magnetic state, resistivity in either direction above 30 K is approximately linear in temperature, shown by the yellow  lines in Fig. \ref{Fig2}(b).

\subsection*{Magnetization and Hall conductivity}

\begin{figure*}[ht!]
\begin{center}
\includegraphics[width=.8\linewidth]{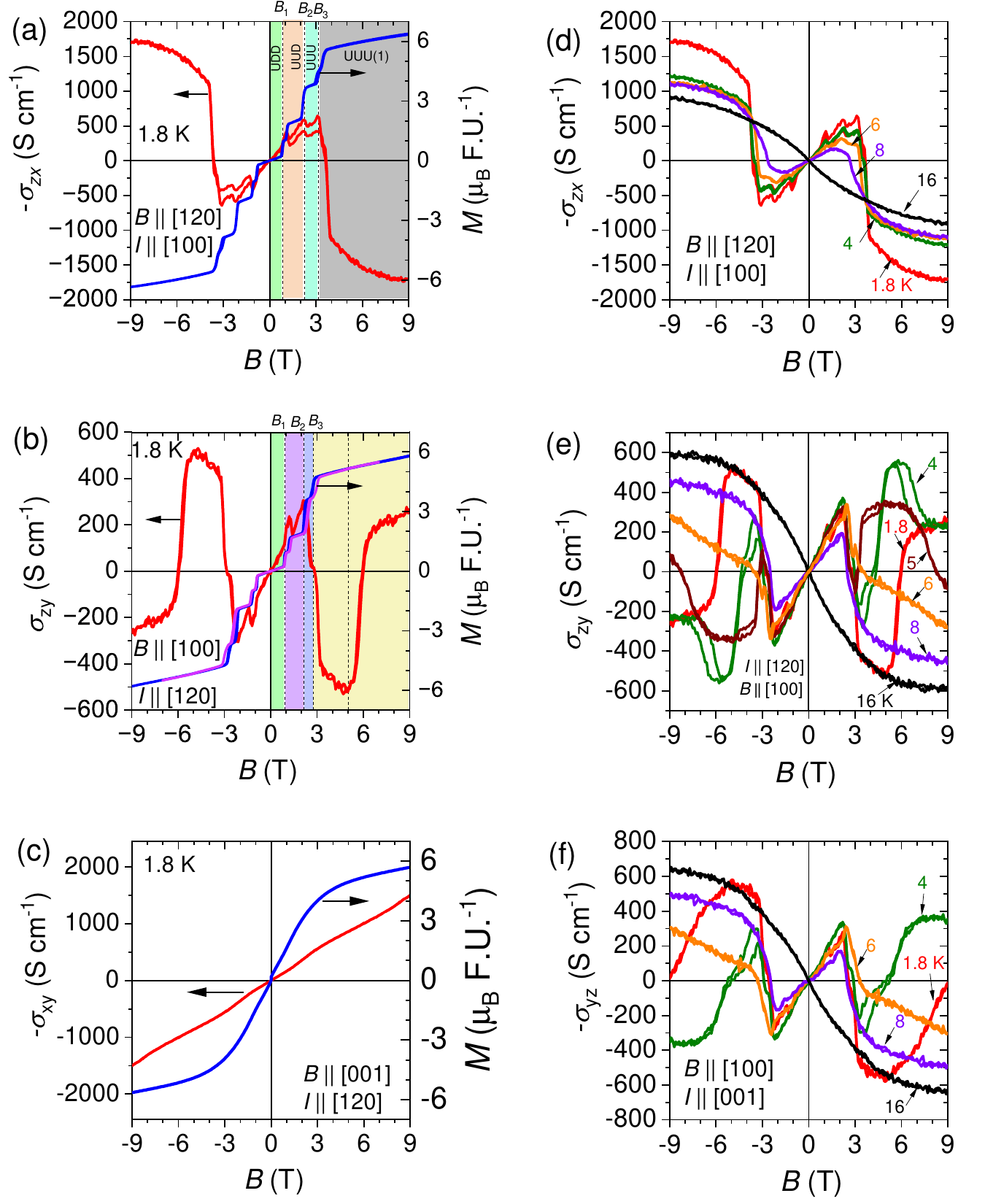}
    \caption{\small \textbf{Magnetization and Hall Conductivity measured with magnetic field $B$ and current $I$ applied along different crystallographic directions.} (\textbf{a}-\textbf{c}) Hall conductivity (red curve, plotted on the left axis) and magnetization (blue curve, plotted on the right axis) as a function of magnetic field for magnetic field $B||[120]$ and $I||[100]$ (\textbf{a}), $B||[100]$ and $I||[120]$ (\textbf{b}), and $B||[001]$ and $I||[120]$ (\textbf{c}). The pink colored curve in panel (\textbf{b}) is the magnetization measured on the Hall bar sample.  (\textbf{d}-\textbf{f}) Hall conductivity as a function of magnetic field at various temperatures for $B||[120]$ and $I||[100]$ (\textbf{d}), $B||[100]$ and $I||[120]$ (\textbf{e}), and $B||[100]$ and $I||[001]$ (\textbf{f}). Dashed lines and shaded regions of different colors serve as guides for the eye, indicating transitions at the magnetic fields $B_1$, $B_2$, and $B_3$ observed in the magnetization measurements. [100], [120], and [001] are defined as $x$, $y$, and $z$ directions as illustrated in the inset of Fig. \ref{Fig2}(b).}
    \label{Fig3}
    \end{center}
\end{figure*}

In this study, we focus on comparing Hall conductivity with magnetization, particularly in the magnetic states  below  $T_{\text{N2}}$. Figure \ref{Fig3}(a) shows the magnetization ($M$, in blue), and Hall conductivity ($-\sigma_{zx}$, in red) for magnetic field ($B=\mu_{\text{0}}H$) $|| [120]$ at 1.8 K. Description of Hall conductivity calculations from the measured transverse and longitudinal Hall resistivities is presented in the Methods section and the corresponding data are presented in Supplementary Figures 6 and 7. Here $-\sigma_{zx}$ is measured with current $I$ along [100] and Hall voltage along [001]. The magnetization curve exhibits three major jumps at the fields $B_1=0.8$ T, $B_2 = 2.1$ T, and  $B_3=3.2$ T, corresponding to metamagnetic transitions from ground state UDD to UUD, UUU, and UUU(1) states, forming the 1/3 plateaus, as depicted in Figs. \ref{Fig1}(c-f). Importantly, it does not fully saturate above $B_3$, but keeps increasing at a sizable rate of approximately 0.15 $\mu_B$/T, suggesting that the assumed Ising MAE is large, but not dominant.

In addition, two smaller jumps in $M$ appear just above $B_1$ and $B_2$ corresponding to 1/6 and 5/6 magnetization plateaus, also observed  previously \cite{zhao2024discrete, zhao2020realization}. Regarding Hall conductivity, $-\sigma_{zx}$ shows a positive slope below $B_3$, with slight changes corresponding to each of the $M$ jumps. At $B_3$, $-\sigma_{zx}$ abruptly changes sign and drops by about 1600 S cm$^{-1}$ before leveling off. Notably, $-\sigma_{zx}(B)$ is not linear like $M$ in the saturated [UUU(1)] state. In the UUD and UUU states, the plateaus in $-\sigma_{xz}$ are clearly split, while those in $M$ are not. This hysteretic behavior in Hall conductivity has been attributed in  Ref. \cite{zhao2020realization} to time-reversal-like metastable toroidal spin-ice structures with opposite chirality, although the microscopic mechanism of coupling toroidicity with the magnetic field is unclear.

When the field is applied along the [100] direction, which is qualitatively different from [120] (see Fig. \ref{Fig3}(b)), three main magnetization steps are again observed at $B_1 = 0.9$ T, $B_2 = 2.1$ T, and $B_3 = 2.9$ T. Notably, between $B=0$ and  $B_2$, the magnetization along [100], and [120] overlaps (see Supplementary Figure 3(a)). However, $B_3 || [100]$ is approximately 0.6 T lower than $B_3 || [120]$, and the magnetization along [100] is about 0.4 $\mu_\mathrm{B}$ per F.U. less than that along [120], consistent with the net $x$ and $y$ projection ratio in the twisted kagome geometry, $2/\sqrt{3}\approx 1.15$. Meanwhile, the differential susceptibility $dM/dT$ above $B_3$ is the same for both the $[100]$ and $[120]$ direction. In Fig. \ref{Fig3}(b), the pink-colored curve represents the magnetization data measured on the same sample used for the Hall resistivity measurements. This was done to rule out any artifact due to the demagnetizing field, which becomes important when comparing magnetization with the Hall conductivity, particularly in this orientation discussed in the next paragraph. The data indicate that the demagnetizing effect between the larger sample used for magnetization measurements and the Hall-bar sample is $\leq$ 0.1 T at 1.8 K.

The Hall conductivity $\sigma_{zy}$ presented in Fig. \ref{Fig3}(b) shows a positive slope below $B_2$, similar to that of $-\sigma_{zx}$, and drops sharply above $B_2$, resembling the behavior of -$\sigma_{zx}$ at $B_3$. While $-\sigma_{zx}$ continues to decrease steadily after the drop at $B_3$, $\sigma_{zy}$ shows a small kink at $B_3$ and completes the sharp drop at 3.1 T. It then increases similarly to $-\sigma_{zx}$ after the drop at $B_3$, up to about $B\approx5$ T, at which point it rises sharply. There is no corresponding feature in magnetization for this sharp rise in $\sigma_{zy}$ (see Supplementary Figure 4 for the derivative of magnetization), highlighting a complete decoupling between Hall conductivity and net magnetization. This  behavior resembles the $-\sigma_{zx}$ plateaus in the UUD and UUU states \cite {zhao2024discrete}, but in this case, the change in $\sigma_{yz}$ without any alteration in magnetization is far more pronounced, pointing to a new and significant magnetism effect on the electronic band structure without any effect on magnetization. To the best of our knowledge, such a drastic change in Hall conductivity without any change in the $M$ response has not been reported in any other materials. As the magnetization (pink curve in Fig.\ref{Fig3}(b)]) and $\rho_{yz}$ were measured on the same sample, any influence of the demagnetizing field on the 5 T upturn observed in $\sigma_{zy}$ can be ruled out. Additionally, the jumps in $\sigma_{zy}$ and $\rho_{zy}$ occur precisely at the same field, further eliminating the possibility of artifacts arising during the conductivity calculations from the measured Hall and longitudinal resistivity (see Supplementary Figure 8).

Finally, when $B$ is applied along [001] direction, the magnetization ($M$) increases steadily, with a very large slope at small fields, $dM/d\mathrm{B}\approx 1.6 $\space$ \mu_{\mathrm{B}}$ F.U.$^{-1}$ T$^{-1}$. Around 3 T the slope gradually changes, reaching a similar slope to the in-plane response only near 9 T. The implications of this behavior, which are inconsistent with the spin-ice model, are presented below in the Discussion section. The Hall conductivity $-\sigma_{xy}$ maintains a positive slope across the entire field range from 0 to 9 T, with only a slight slope change near 3 T as depicted in Fig. \ref{Fig3}(c). It is worth noting that the slope magnitude of $-\sigma_{zx}$, $\sigma_{zy}$, and $-\sigma_{xy}$ below 2 T remain fairly consistent, ranging from $\sim 120$ to 160 S cm$^{-1}$ T$^{-1}$, suggesting that the apparent Hall carrier concentration remains nearly constant in these cases. 

To get a deeper insight into the various features observed, particularly in $-\sigma_{zx}$ and, $\sigma_{zy}$, we analyzed their temperature dependence for different directions of $I$ and $B$. Figure \ref{Fig3}(d) illustrates the behavior of $-\sigma_{zx}$ as the temperature decreases from 16 K to 1.8 K. At 16 K, $-\sigma_{zx}$ exhibits an overall negative slope, indicating that electrons are the dominant carriers. Just below $T_{\text{N1}}$, $-\sigma_{zx}$ exhibits a slight positive slope, suggesting that holes have become the primary charge carriers, as seen in the 8 K data (extended data with additional temperatures are provided in Supplementary Figure 5). However, at $B_3$, $-\sigma_{zx}$ drops sharply, and returns to a negative slope, similar to the behavior observed at 16 K data. This trend in $-\sigma_{zx}$ continues as the temperature decreases to 1.8 K. Above $B_3$, the magnitude of $\sigma_{zx}$ increases progressively as the temperature decreases from $T_{\text{N2}}$ to 4 K. Nevertheless, a significant increase in this magnitude is observed between 4 K and 1.8 K, indicating the emergence of an additional Hall signal at 1.8 K.

The temperature dependence of $\sigma_{zy}$ is depicted in Fig. \ref{Fig3}(e). After the $B_2$ drop, the magnitude of $\sigma_{zy}$ at 1.8 K does not exceed its value at 16 K, which corresponds to the non-magnetic state. Between $B_3$ and 5 T at 1.8 K, $\sigma_{zy}$ tracks the 16 K data. However, above 5 T, the 1.8 K data turns sharply followed by a curving off. This upturn and subsequent curving off are observed on warming up to 5 K, but disappear at 6 K. These upturns and the subsequent downturns observed at 4 and 5 K do not have any corresponding features in the magnetization (see Supplementary Figs. 4 and 9) as in the case of 1.8 K-data depicted Fig. \ref{Fig3}(b). Here, the high-field downturn is not observed at 1.8 K. However, upon performing the measurement up to 14 T (measured in another sample for reverification), we observed the downturn at 1.8 K at around 13 T (see Supplementary Figure 10). Above 6 K, $\sigma_{zy}$ shows a monotonic decrease with the magnetic field after the initial drop. Notably, $\sigma_{zy}$ does not display hysteresis at 1.8 K, although significant hysteresis is observed at 4 K during the initial upturn and the subsequent downturn, centered at around 3.2 and 7 T, respectively. At 5 K, the hysteresis is observed only after the first upturn. Similar Hall conductivity behavior is observed when current and voltage directions are reversed, with the magnetic field aligned along [100] $-\sigma_{yz}$) as depicted in Fig. \ref{Fig3}(f). The behavior below $B_2$ remains consistent, but above $B_2$, the sharp slope changes in $-\sigma_{yz}$ shift slightly toward higher $B$.
 
In summary, Hall conductivity measurements reveal four notable features: (1) The Hall conductivity, which shows electron-like behavior in the non-magnetic state, shifts to hole-like behavior with magnetic ordering below $B_2$ and $B_3$ for magnetic fields aligned along the [100] and [120] directions, respectively. (2) At $B_2||[100]$ and $B_3||[120]$, there is an abrupt sign change in Hall conductivity, reverting to electron-like behavior. (3) A significant enhancement in Hall conductivity is observed immediately after the $B_3$ sign change, particularly below 4 K, when B is applied along the [120] direction. (4) Below $T_{N2}$, Hall conductivity below 5 K measured with B along [100] exhibits multiple changes after the $B_2$ drop without any corresponding change in magnetization.

\subsection*{Electronic band structure calculations}

\begin{figure*}[ht!]
\begin{center}
\includegraphics[width=.8\linewidth]{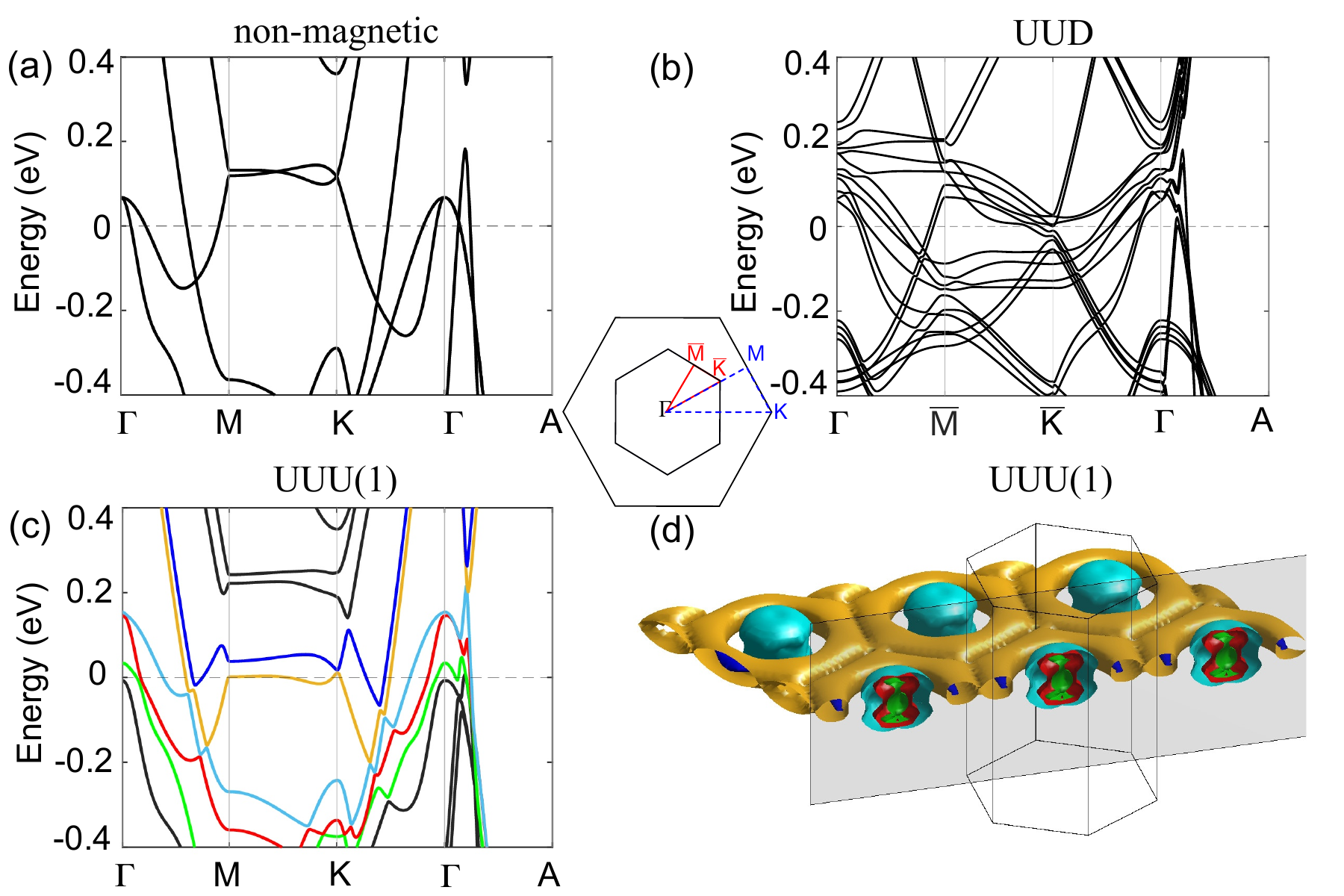}
    \caption{\small \textbf{Calculated electronic structure of HoAgGe} (\textbf{a}) Band structure in the non-magnetic state. (\textbf{b}) Electronic structure in $\sqrt{3}\times\sqrt{3}$ magnetic state UUD. (\textbf{c}) Electronic structure in $1\times1$ [UUU(1)] stae. (\textbf{d}) Fermi surface in the UUU(1) state. In the nonmagnetic state  Fermi surface is very similar, with the main difference being that the orange donuts are disconnected, though they retain the electron character. The colors correspond to those in panel (\textbf{c}).}
    \label{Fig4}
    \end{center}
\end{figure*}

To understand the microscopic origins of these Hall conductivity features, we performed electronic band structure calculations, as depicted in Fig. \ref{Fig4}. These calculations were conducted in three states: the non-magnetic state, the magnetic UUD state, and the UUU(1) state. As shown in Figs. 1(c-e), the UDD, UUD, and UUU states share the same $\sqrt{3}\times\sqrt{3}$ magnetic unit cell. However, in the UUU(1) state, the magnetic unit cell is reduced to $1\times1$, the same as in the non-magnetic state, but with ordered spins. 

The band structure in the non-magnetic state, shown in Fig. \ref{Fig4}(a), reveals a large electron pocket along the hexagonal first Brillouin zone boundary,
and two smaller hole pockets located along $\Gamma-A$. The  electron pockets generate the electron-like Hall conductivity, especially for in-plane measurements.  Fig. \ref{Fig4}(a) shows, $\sim 150$ meV above the Fermi energy, these quasi-2D bands hardly changing along $MK$ but highly dispersive along $M-\Gamma$, and $K-\Gamma$, forming a Dirac-like crossing at $K$. 

With the onset of antiferromagnetic ordering, in the UUU state, as shown in Fig. \ref{Fig4}(b), the band structure becomes considerably more complex due to band folding from the tripled unit cell. The individual pockets, first of all, the electron donuts in Fig. \ref{Fig4}(d), reconnect and the topology of the Fermi surface changes dramatically --- even though the band's dispersion changes only moderately. Experimental evidence of hole-like conductivity in this state suggests that the reconnection of the electron pockets makes them hole-like, which, although challenging to quantify, is reflected in the band structure. The small spikes observed in Hall conductivity at $B_1$ and $B_2$ correspond to spin flips at these magnetic fields. However, these spin flips do not significantly alter the magnetic unit cell, resulting in minimal changes to the sign and magnitude of the Hall conductivity. This is further supported by the overall positive slope of $-\sigma_{xy}$ in Fig. \ref{Fig3}(c) where the magnetic unit cell does not change.

In the UUU(1) state, the magnetic unit cell reduces to $1\times 1$, eliminating band folding and resulting in a Fermi surface similar to the non-magnetic state, shown in Fig. \ref{Fig4}(c). The magnetic ordering induces  spin splitting, pushing the quasi-2D bands and Dirac-like crossings to the Fermi energy, which makes the orange donuts touch but does not change their character, restoring the electron-like Hall conductivity. Consequently, above $B_3$, where the UUU(1) magnetic structure stabilizes, the sign of the Hall conductivity changes.

A more rigorous verification of the calculated band structure, particularly the Fermi surface, comes from the longitudinal transport. We calculated the plasma frequency in the paramagnetic state and found that $\omega_{pz}^2\approx 3.2\omega_{px}^2$. Using the basic Drude model, where $\sigma_\alpha=\omega_{p\alpha}^2\tau_\alpha/4\pi$ (with $\tau_\alpha$ representing the anisotropic relaxation time), we observed that the resistivity shown in Fig. \ref{Fig2}, is almost identical for the [120] and [100] directions, as expected. Additionally, above $\sim 30$ K, it follows a perfectly linear trend, indicating that the primary scattering agents are excitations with energy $\alt 4\times30 k_B\sim 10$ meV, too low for phonons; the best candidate for this role are, obviously, spin fluctuations. This agrees with the exchange coupling estimate from Ref. \cite{zhao2020realization}, and our first-principles calculations presented in detail in the Supplementary Section 1. The resistivity in this range is well described by $\rho_x(T)$=70.6+0.82 T $\mu\Omega$ cm, $\rho_z(T)$=30.1+0.25 T $\mu\Omega$ cm (see Fig. \ref{Fig2}(b)). The ratio of the slopes is 3.28, in excellent agreement with DFT, confirming the accuracy of the calculated Fermi surface. Note that the ratio of the constant terms above is smaller, 2.34, indicating that  scattering off defects is anisotropic, with $\tau_x$ about 30\% smaller than $\tau_z$.

\section*{Discussion}

The experimentally reported magnetic structures for $B||[120]$ account for most of the observed features in the Hall conductivity $-\sigma_{zx}$, including small spikes during transitions into the UUD and UUU phases, and the abrupt sign reversal in the UUU(1) phase. While these changes in the transport and magnetic properties are clearly associated with discontinuous spin reorientations, suggested in Ref. \cite{zhao2020realization}, several qualitative effects remain unexplained /cannot be explained by the ideal spin-ice model with infinite MAE. Thus, obtaining experimental estimates of both the exchange coupling and the MAE is crucial. To this end, we analyzed magnetization data above $B_3$ for in-plane fields and over the full range for out-of-plane fields. The details are provided in the Supplementary Section 2, where we used slopes and magnetization values in all three directions to find that the in-plane magnetic anisotropy can be described by the lowest-order term $K_{||}\cos^2\phi$ (with $\phi$ representing the deviation from the easy axis), and $K_{||}\approx -5.6\pm0.6$ meV, substantial  compared to the exchange coupling but not overwhelmingly larger. 

Interestingly, the out-of-plane anisotropy cannot be described as $K_\perp \cos^2\theta$ ($\theta$ characterizes tilting away from the plane). Instead, it requires higher order terms such as  $K'_\perp \cos^4\theta$ and $K_\perp'' \cos^6\theta$, as seen in similar studies on $R$Mn$_6$Sn$_6$ compounds \cite{lee2023interplay}. Moreover, the lowest-order term for out-of-plane anisotropy is anomalously soft, with $K_\perp\approx1$ meV, suggesting very strong spin-canting fluctuations. Thus, the ideal spin-ice model is insufficient, and the unexplained features seen in this and previous studies likely arise from deviations from this model.

For $B||[100]$, the magnetic structure remains unknown. However, the Hall conductivity in this configuration suggests that, similar to $B||[120]$, a change in the magnetic unit cell leads to a sign reversal of $\sigma_{zy}$. The sign change of $\sigma_{zy}$ in the UUU(1) phase, where magnetization increases linearly with $B$, is unexpected and intriguing. We tentatively attribute it to ordering processes in the Ho$_1$ sublattice. In the limit  $K\gg MB_x$ ($K$ representing the in-plane anisotropy and $M$ the Ho moment), the Ho$_1$ spins are perpendicular to $x$ and do not contribute to the total magnetization. In this limit, these spins do not couple either to the external field $B_x$ or to the saturated Ho$_{2,3}$ sublattices. Instead, they are coupled to each other through weak third-neighbor interaction, potentially mediated by Dzyaloshinskii-Moriya interaction (DMI) (for  Ho$_1$--Ho$_1$ bonds, the DMI vector $\mathbf{D}||z$), which couples the $x$ and $y$ projections. Along with the finite stiffness of Ho spins with respect to canting away from the easy directions, this suggests that twisting the kagome lattice introduces a unique mechanism for coupling spins to the transport, driven by the resulting frustration. The  irreversible Hall plateaus with the same magnetization, as observed in Ref. \cite{zhao2020realization}, as well as the sign-flip of the Hall conductivity without a corresponding change in magnetization seen in our work, may also stem from deviations from the ideal Ising spin dynamics and the spin-ice rule. Our additional calculations, which will be the subject of a future publication, have, in fact, successfully reproduced all magnetization features, including the one-third jumps and the two secondary intermediate jumps, resulting in a rich and intriguing spin-ice Heisenberg model.

Here, we clarify the novelty of our findings, as Ref. \cite{roychowdhury2024enhancement} also presents Hall resistivity measurements for $B || [100]$. However, their data (Supplementary Figure 5 in Ref. \cite{roychowdhury2024enhancement}) differs significantly from our results shown in Supplementary Figure 7(d). In fact, their data closely resembles our   $-\rho_{yz}$ measured with $B || [120]$, $I || [001]$, as presented in Supplementary Figure 11. Examining the Laue data in their Supplementary Figure 1, it is evident that the applied magnetic field direction in their measurements is apparently [120], not [100]. Therefore, our $B || [100]$ data should not be directly compared to that in Ref. \cite{roychowdhury2024enhancement}. Furthermore, we conclude that for $B ||[120]$, the abrupt sign change in Hall conductivity (or resistivity) below 6 K is due to a change in carrier concentration caused by the modification of the magnetic structure. This feature has been attributed to the topological Hall effect in Ref. \cite{roychowdhury2024enhancement} (Fig. 4D).

\section*{Conclusion}

The kagome lattice has been extensively studied for its magnetic frustration, and electronic topological properties, including flat and Dirac bands. However, the role of structural twisting in this lattice has been largely overlooked. In this regard, our findings on HoAgGe open a new pathway for investigating the interaction between structural distortion, magnetic frustration, and their influence on electronic topology. This is  particularly intriguing as spin-polarized quasi-2D bands and Dirac crossings/anti-crossings are in close proximity of the Fermi energy. Additionally, HoAgGe is a part of the larger $RTX$ family of compounds \cite{merlo1996rmx,iandelli1985structure,morosan2005angular,baran2000magnetic,gondek2007magnetic,katoh2004magnetic} (where $R$ is a rare earth element, $T$ is a transition metal, and $X$ is either Ge or Si), providing the ability to tune magnetism through different combinations of $R$, $T$, and $X$ atoms.

\section*{Methods}\label{S2}

\textbf{Crystal growth and structural characterization.} Single crystals of HoAgGe were grown by the flux method considering eutectic point of Ag-Ge. Ho pieces (Thermo Scientific 99.9\%), Ag shots (Thermo Scientific 99.999\%), and Ge pieces (Thermo Scientific 99.9999\%)  were loaded into a 2 ml aluminum oxide crucible in a molar ratio of 1:7.6:2.5. The crucible was then sealed in a fused silica ampule under vacuum. The sealed ampule was heated to 1175 $^{\circ}$C over 10 hours, kept at 1175 $^{\circ}$C for 10 hours, and then cooled to 825 $^{\circ}$C over one week. Once the furnace reached 825 $^{\circ}$C, the tube was centrifuged to separate the crystals in the crucible from the molten flux.  Several well-faceted long crystals [see the inset in Fig. \ref{Fig2}(b) for an optical image of a polished crystal] up to 150 mg were obtained in the crucible. The crystal structure was verified by Rietveld refinement \cite{Mccusker1999} of a powder X-ray diffraction pattern collected on a pulverized single crystal using a Rigaku Miniflex diffractometer. The Rietveld refinement was performed using the FULLPROF software \cite{Rodriguez-carvajal1993} and is depicted in Supplementary Figure 12. Selected data obtained from the Rietveld refinement are presented in Supplementary Table 2. Crystal and magnetic structures were drawn using VESTA software \cite{vesta}. Before each measurement, every crystal was oriented along the  [001], [100], and [120] directions using both an X-ray Laue diffractometer and an X-ray powder diffractometer. The powder diffractometer was used to confirm the [001], [120] and [110] Bragg peaks after the final polishing.

\textbf{Magnetic property measurements.} Most of the DC magnetic susceptibility and magnetization data presented in the manuscript and Supplementary Information were measured using a Quantum Design (QD) Dynacool Physical Property Measurement System (PPMS).  The AC Measurement System (ACMS II) option was used for the DC magnetization measurements. The magnetization data presented in Supplementary Figure 10 was measured using the Vibrating Sample Magnetometer(VSM) option in a QD PPMS equipped with a 14 T magnet. The magnetization measurement on the Hall bar sample, presented in Fig. \ref{Fig3}(b) and Supplementary Figure 9, was conducted using a QD Magnetic Property Measurement System -3 (MPMS-3) with a 7 T magnet. This measurement was performed after removing the electrical contacts from the sample previously used for $\rho_{zy}$ measurements and carefully removing the epoxy used to attach the electrical contacts.

{\textbf{Resistivity and magnetotransport measurements.} Resistivity and Hall resistance were measured using the resistivity option of a QD DynaCool PPMS, equipped with either a 9 T or 14 T magnet. All measurements were performed using the four-probe method in a Hall bar geometry. Electrical contacts were made with 25 $\mu$m platinum wires, secured using Epotek H20E silver epoxy, resulting in typical contact resistances below 30 $\Omega$. An excitation current of 4 mA was applied for all electrical transport measurements.

The samples were polished to dimensions of $\sim 1.00 \times 0.40 \times 0.15$ mm, with the long axis oriented along either the [100], [120] or [001] crystallographic direction.

The longitudinal resistivity, $\rho_{ii}$, was calculated from the measured longitudinal resistance, $R_{ii}$, using the relation: $\rho_{ii} =R_{ii}A/l$, where $A = td$ is the cross-sectional area ($t$ is the thickness, and $d$ is the width of the sample), and $l$ is the length of the sample between the two voltage contacts.

The Hall resistivity  was calculated using the relation: $\rho_{ij}$ = $R_{ij}t$, where $R_{ij}$ = $V_i$/$I_j$ and $t$ represent the measured Hall resistance and the sample thickness, respectively. Here, $V_i$ is the transverse voltage developed along the $i$-direction in the presence of a magnetic field along $k$, with current $I_j$ in the $j$- direction. The indices $i$, $j$, and $k$ are mutually orthogonal, representing $x$, $y$ or $z$ directions in different measurement geometries.

To account for the contact misalignment, the antisymmetric (Hall) and symmetric (magnetoresistance) contributions to the resistivity and Hall data, respectively were corrected using symmetrization and antisymmetrization techniques as follows:
\begin{equation*}
  \rho_{ii}=\frac{\rho_{ii}(+B)+\rho_{ii}(-B)}{2},~~   \rho_{ij}=\frac{\rho_{ij}(+B)-\rho_{ij}(-B)}{2}.
\end{equation*}

The longitudinal resistivity and Hall resistivity data were acquired in a four-loop sequence: the magnetic field $B$ was swept from +$B_{\text{max}}$ to -$B_{\text{max}}$, and then back from -$B_{\text{max}}$ to $+B_{\text{max}}$. A similar protocol was used for $M$ vs $B$ measurements. Symmetrization and antisymmetrization were applied to data obtained during $B$-field sweeps from $+B_{\text{max}}$ to 0 T and from $-B_{\text{max}}$ to 0 T, as well as from 0 T to $+B_{\text{max}}$ and from 0 T to $-B_{\text{max}}$.

Hall conductivity for various directions was calculated using the following relations:  
\begin{equation*}
\sigma_{xy} = \frac{-\rho_{xy}}{\rho_{xx}^{2}}, \sigma_{zx} = \frac{-\rho_{zx}}{\rho_{xx}\rho_{zz}}, \sigma_{yz} = \frac{-\rho_{yz}}{\rho_{yy}\rho_{zz}}, \sigma_{zy} = \frac{-\rho_{zy}}{\rho_{zz}\rho_{yy}}.
\end{equation*}
Here, $\rho_{xx}$, $\rho_{yy}$, and $\rho_{zz}$ represent the longitudinal resistivities for current along the [100], [120], and [001] crystallographic directions, respectively, with  the magnetic field applied perpendicularly. The Hall and longitudinal resistivity terms $\rho_{yz}$ and $\rho_{zz}$, as well as $\rho_{zy}$ and $\rho_{yy}$ were measured using two different crystals from the same growth batch and used to calculate $\sigma_{yz}$ and $\sigma_{zy}$. The terms $\rho_{xx}\rho_{yy}$  and $\rho_{yy}\rho_{zz}$ in the denominators of $\sigma_{xz}$, $\sigma_{yz}$ and $\sigma_{zy}$ account for the resistivity anisotropy in the hexagonal system, as shown in Fig. \ref{Fig2}. For $\sigma_{xy}$, $\rho_{xx}$ = $\rho_{yy}$ is used due to the isotropic in-plane resistivity. The conductivity relations used here are valid under the condition: $\rho_{ii}\rho_{jj}$ $\gg$ $\rho_{ij}\rho_{ji}$.

\textbf{Electronic structure calculations.} Electronic structure calculations were performed using Vienna ab initio Simulation Package (VASP) \cite{kresseEfficientIterativeSchemes1996} within projector augmented wave (PAW) method \cite{blochlProjectorAugmentedwaveMethod1994}. The Perdew-Burke-Enzerhof (PBE) \cite{perdewGeneralizedGradientApproximation1996} generalized gradient approximation was employed to describe exchange-correlation effects. For the ordered states, we added a Hubbard U correction with the fully localized limit double-counting recipe \cite{liechtensteinDensityfunctionalTheoryStrong1995,dudarevElectronenergylossSpectraStructural1998}, to account for the strongly correlated Ho $4f$ states and their localized magnetic moments. The effective parameter $U-J=8$ eV was used.

It is well-known that in such case when f-states are far removed from the Fermi level the main challenge in DFT is to reproduce the correct, according to the Hund's rules, ground state, and when this condition is met the results are very reliable\cite{lee2023interplay}. In our case, we verified that
the orbital moment on the Ho site obtained from the calculations is consistently 6 $\mu_{\mathrm{B}}$ which satisfies Hund's rules and ensures that the electronic structure near the Fermi level is reliable. For the paramagnetic state, the open-core approximation is employed, where the $4f$ electrons in the Ho pseudopotential are treated as part of the frozen core, simulating the average effect of the disordered magnetic moments. The plasma frequencies are obtained by integration of the Fermi velocity implemented in VASP via the LOPTICS tag \cite{loptic}. 

\begin{acknowledgments}
This work was primarily supported by the U.S. Department of Energy, Office of Science, Basic Energy Sciences, Materials Science and Engineering Division. IM was supported by the Office of Naval Research through grant \#N00014-23-1-2480.
\end{acknowledgments}

\section*{Author contributions}
H.B. and N.J.G. conceived the idea and coordinated the project. H.B. grew single crystals. R.R. helped in the crystal growth. H.B. performed magnetic and magnetotransport measurements. B.G.M., L.F. and J.F.M. helped in some magnetotransport measurements. H.B. carried out the data analysis. I.I.M. and  P.H.C. contributed to the DFT calculations. H.B. and N.J.G. wrote the manuscript with input from I.I.M. and P.H.C. All authors contributed to the discussion of the results.

\widetext
\begin{center}
\pagebreak
\hspace{0pt}
\vfill
\textbf{\large Supplementary Material}
\vfill
\hspace{0pt}
\end{center}
\FloatBarrier

\setcounter{equation}{0}
\setcounter{figure}{0}
\setcounter{table}{0}
\setcounter{page}{1}
\makeatletter
\renewcommand\thesection{S\arabic{section}}
\renewcommand{\theequation}{S\arabic{equation}}
\renewcommand{\thetable}{S\arabic{table}}
\renewcommand\thefigure{S\arabic{figure}}
\renewcommand{\theHtable}{S\thetable}
\renewcommand{\theHfigure}{S\thefigure}

\begin{figure*}[ht!]
\begin{center}
\includegraphics[width=.6\linewidth]{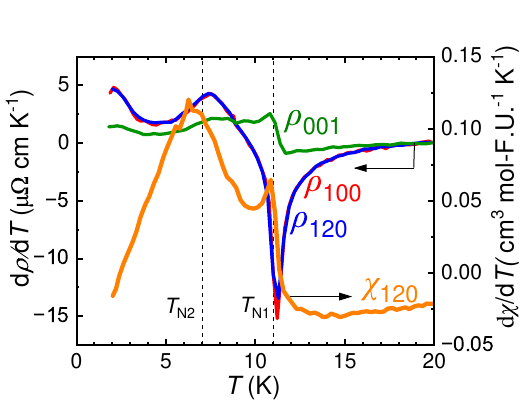}
    \caption{\small \textbf{Derivative of Resistivity and Susceptibility.} The temperature derivative of resistivity of HoAgGe is plotted as a function of temperature for current ($I$) applied along the [100], [120] and [001] directions (left axis). The temperature derivative of susceptibility ($\chi_{120}$) is shown on the right axis. $T_{\text{N1}}$ and $T_{\text{N2}}$ denote the two N\'eel temperatures. Dashed lines are included as a visual guide.}
    \label{FigS1}
    \end{center}
\end{figure*}

\section{E\lowercase{xchange} c\lowercase{oupling} c\lowercase{onstants}}

A numerical-orbital-based \cite{ozakiVariationallyOptimizedAtomic2003}
DFT code OpenMX \cite{OpenMX} was used for the calculations of magnetic
properties. The exchange coupling constants up to 5$^{th}$ nearest neighbors (defined in Supplementary Figure \ref{jnn}) are calculated perturbatively using Green\textquoteright s function method \cite{antropovExchangeInteractionsMagnets1997,katsnelsonFirstprinciplesCalculationsMagnetic2000}
implemented in OpenMX 3.9 \cite{terasawaEfficientAlgorithmBased2019,omxgf2004}. 
The effective spin-Hamiltonian is defined as 
\begin{equation}
    \begin{split}
    H & =\sum_{\left\langle ij\right\rangle _{1}}J_{1}\hat{n}_{i}\hat{n}_{j}+\sum_{\left\langle ij\right\rangle _{2}}J_{2}\hat{n}_{i}\hat{n}_{j}+\sum_{\left\langle ij\right\rangle _{3a}}J_{3a}\hat{n}_{i}\hat{n}_{j}+\sum_{\left\langle ij\right\rangle _{3b}}J_{3b}\hat{n}_{i}\hat{n}_{j}\\
     & +\sum_{\left\langle ij\right\rangle _{4}}J_{4}\hat{n}_{i}\hat{n}_{j}+\sum_{\left\langle ij\right\rangle _{5}}J_{5}\hat{n}_{i}\hat{n}_{j},
    \end{split}\label{sham}
\end{equation}
where $\hat{n}$ is the unit vector along local easy-axis, $J_n$ is the exchange coupling parameters for the n-th NN interaction summarized in Table \ref{T1}, and summation is over all bonds of a given type. Note that there are two inequivalent 3rd NN interactions denoted as $J_{3a}$ and $J_{3b}$.

\begin{figure*}[ht!]
\begin{center}
\includegraphics[width=.5\linewidth]{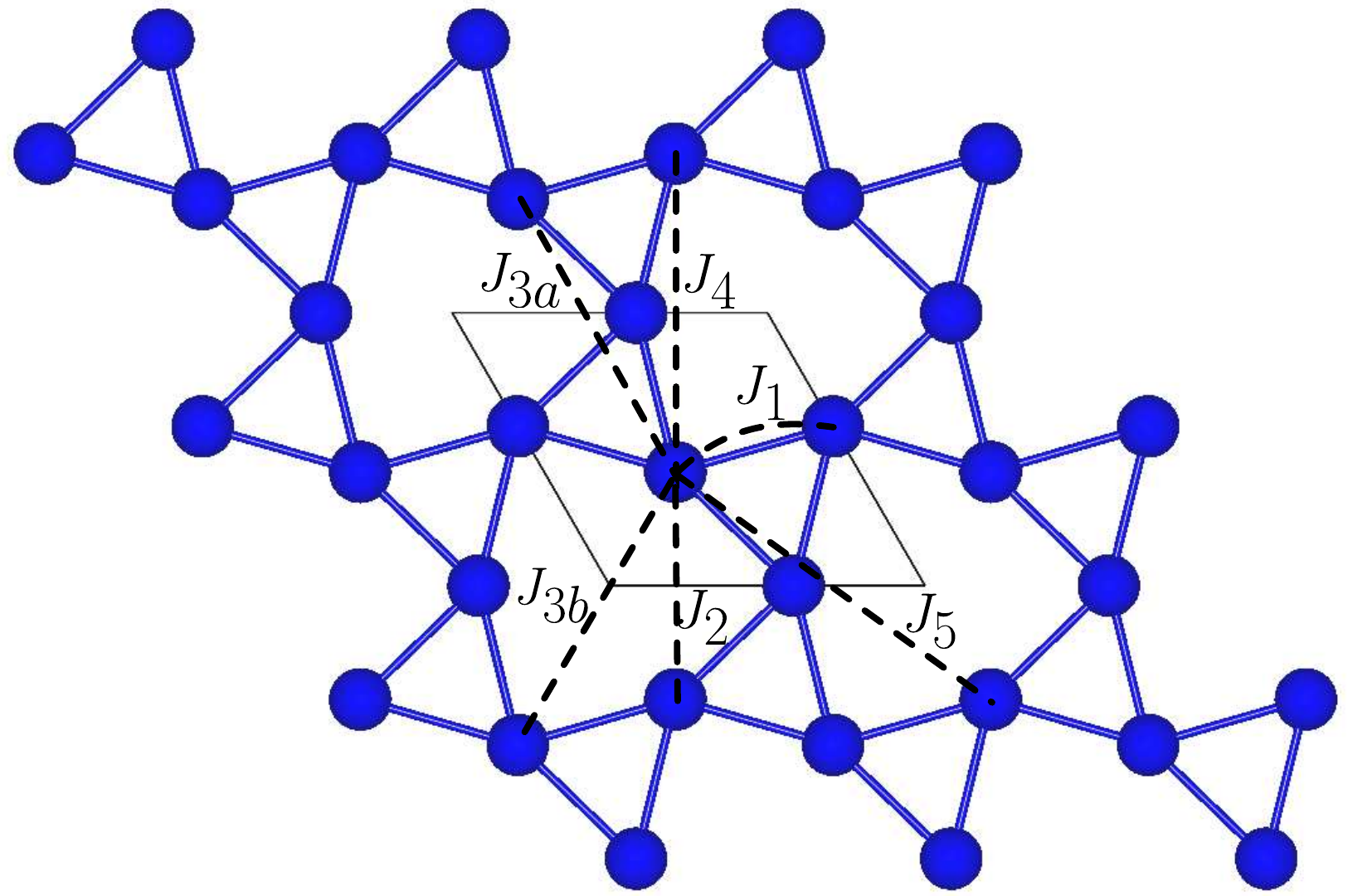}
    \caption{\small \textbf{Exchange coupling constants.} Definitions of effective exchange coupling constants of HoAgGe up to 5 nearest neighbor.}
    \label{FigS2}
    \end{center}
\end{figure*}

\begin{table}[h]
\begin{center}
\caption{Effective exchange coupling constants up to 5th nearest neighbor calculated using Green's function method.}\label{T1}
    \begin{tabular}[c]{lcc} 
       \hline 
        & \# of NN & meV\\
        \hline 
        $J_{1}$  & 4 &  3.154\\
        $J_{2}$  & 2 & -1.395\\
        $J_{3a}$ & 4 & -0.412\\
        $J_{3b}$ & 2 & -0.552\\
        $J_{4}$  & 2 & -0.073\\
        $J_{5}$  & 4 & -0.212\\
      \hline 
    \end{tabular}
   \end{center}
\end{table}


\section{M\lowercase{agnetization}}

\begin{figure*}[ht!]
\begin{center}
\includegraphics[width=.6\linewidth]{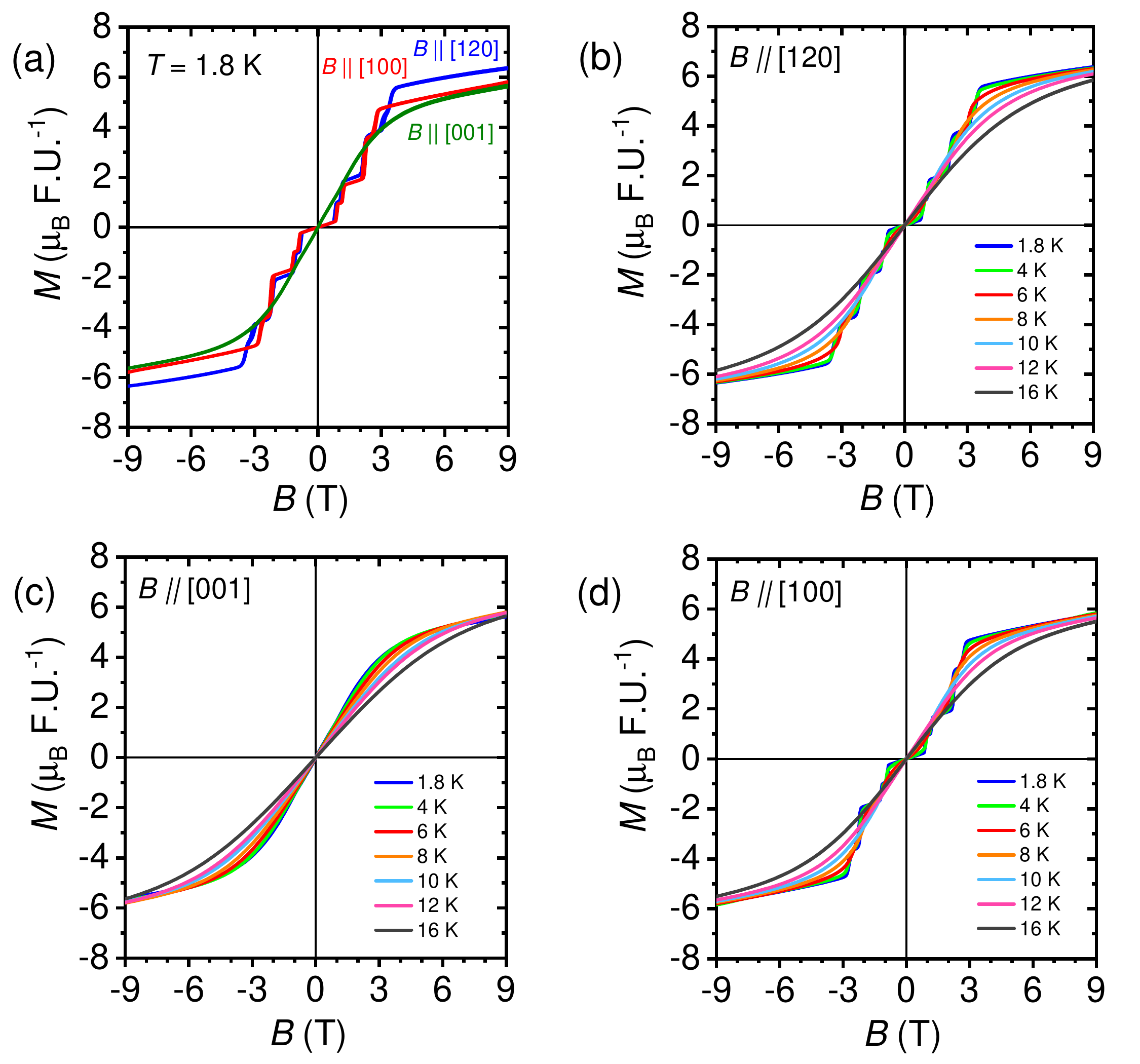}
    \caption{\small \textbf{Magnetization vs. magnetic field.} a) Magnetization ($M$) of HoAgGe as a function of magnetic field ($B$) at 1.8 K is plotted for $B$ applied along the three crystallographic directions: [100] (red), [120] (blue), and [001] (green). b-d) $M$ vs. $B$ is shown for temperatures between 1.8 and 16 K for $B||[120]$ (b), $B||[001]$ (c), and $B||[100]$(d).}
    \label{FigS3}
    \end{center}
\end{figure*}

\begin{figure*}[ht!]
\begin{center}
\includegraphics[width=.4\linewidth]{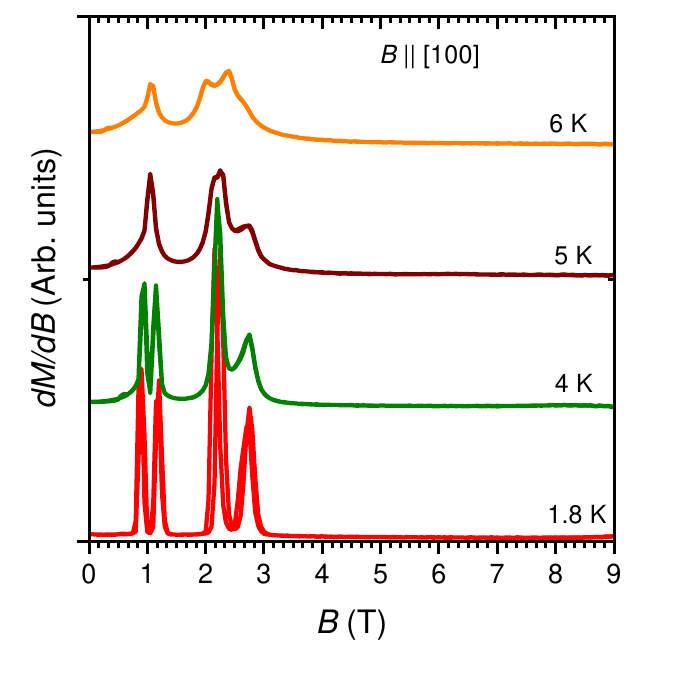}
    \caption{\small \textbf{Derivative of Magnetization.} The derivative of magnetization of HoAgGe with respect to the magnetic field, measured as a function of the magnetic field between 1.8 and 6 K for $B||[100]$. There is no magnetization jump or slope change above 3 T at 1.8 K, nor at the corresponding fields for other temperatures.}
    \label{FigS4}
    \end{center}
\end{figure*}

Magnetization, $M$, for the case of  $B||[120]$ is both rich and intriguing. To fully understand the interactions within the system, it is necessary to consider all three spatial directions. Previous studies \cite{zhao2024discrete-s, zhao2020realization-s} explained these phenomena using a ``spin-ice'' model, which assumes infinite easy-axis anisotropy, with the axis orientation varying from site to site according to the three-fold symmetry. However, the non-zero slope of $M(B)$ in the fully saturated spin-ice phase suggests that even at magnetic fields of about 5 T, the Ho spins are still slightly canted. This is consistent with the nonlinear Hall conductivity observed for a linear increase in magnetization (see Figure 3(a) and Supplementary Figure \ref{FigS2}(a)). Using this information and assuming a lowest-order in-plane angular anisotropy, we can express the total anisotropy energy for an individual site as $K_{||}M^2\cos^2{(\phi-\alpha)}$, where $\phi-\alpha$ is the canting angle away from the easy axis, which is assumed to be at an angle $\alpha$ to the Cartesian $x$ ([100]). Zeeman energy will then be $-hM$; here $M$ is the full moment of an individual Ho ion, and $h$ the applied field. Note that for Ho$_1$, $\alpha=90^\circ$, and for Ho$_{2,3}$, $\alpha=\pm 30^\circ$. Neglecting the smaller exchange coupling, we find: $\chi_{120}$ = [$\chi_{120}$(Ho$_2$)+$\chi_{120}$(Ho$_3$)]/3 = 1/4K and the net magnetic moment per site: $M_{120}(B)=(2/3) M+h/4K$. Extracting these parameters from the experimental data in the saturated phase, we find $M\approx 7.7$ $\mu_B$, in excellent agreement with neutron data \cite{zhao2024discrete-s}, and $KM^2\approx 6.2$ meV, a typical value for 4f ions. 

Similarly, for $B||[100]$, $M_{100}(B)$=(1/$\sqrt{3}) M+h/4K$. Here, all three ions contribute to $\chi$. By fitting our data, we find  $M\approx 7.5$ $\mu_B$, $K M^2\approx 5.0$ meV, confirming the consistency of the model (within $\pm$ 10\%). This analysis shows that while the spin-ice (infinite anisotropy) model is a useful first approximation, it remains a rough one.

On the other hand, the experimentally measured $M_{001}(B)$ does not align with the spin-ice model at all (see Supplementary Figure \ref{FigS2}(a)). In this case, assuming a $\cos^2\phi$ easy-plane anisotropy, $\chi=1/2K$. The low-field slope of the experimental data is at least ten times higher than the saturation-regime in-plane slope, suggesting that the quadratic anisotropy coefficient $K_2M^2\approx 1.0$ meV. Moreover, the slope gradually decreases with increasing field, dropping to about $0.13$, even lower than the in-plane values. The ramification is that the out-of-plane anisotropy is strongly non-quadratic. In f-electron ions, three terms ($\cos^2\phi$, $\cos^4\phi$, and $\cos^6\phi$) are permitted, without any parametric smallness\cite{lee2023interplay-s}. Indeed, the nonlinear susceptibility is reasonably well described by the 6-th order Hamiltonian, $E=K_2M^2\cos^2\phi+K_4M^4\cos^4\phi+K_6M^6\cos^6\phi$, where $K_2M^2$ is as above, $K_4/K_2=-0.4$, and $K_6/K_2=0.11$. This indicates that the out-of-plane spin dynamics differ significantly from the in-plane dynamics and are far from the strong anisotropy limit of the spin-ice model. Notably, if the quartic coefficient were 60\% larger, the system would transition abruptly to an easy-cone anisotropy with the cone angle approximately 30$^{\circ}$ from the plane. This effect could occur if Ho were substituted by another rare-earth element, as seen in the 166 family of materials\cite{lee2023interplay-s}.

\section{M\lowercase{agnetotransport}}

The extended Hall conductivity data presented in Figure 3 of the manuscript is also presented in Supplementary Figure \ref{FigS3}. The longitudinal resistivity and Hall resistivity versus magnetic field data, used to calculate the conductivity in Supplementary Figure \ref{FigS3}, are presented in Supplementary Figures \ref{FigS5} and \ref{FigS4}, respectively. Additionally, two new samples from a different growth batch were prepared to reverify the Hall conductivity behavior along $B||$[100]. Magnetization ($M$), Hall resistivity ($\rho_{zy}$) and longitudinal resistivity ($\rho_{yy}$) as functions of the magnetic field were measured on the same sample.

Initially, the magnetization was measured on a polished and oriented single crystal. The crystal was then trimmed into a Hall bar geometry to measure $\sigma_{zy}$ and $\rho_{yy}$. These measurements were conducted in magnetic fields of 14 T, as shown in Supplementary Figure \ref{FigS14T}. All features observed in the previously measured samples, as presented in Figures 2 and 3 of the manuscript, were reproduced in these new samples, including the consistent observation of magnetization steps.

The upturn of $\sigma_{zy}$ (or in $-\rho_{zy}$) following magnetic saturation in this new sample [Supplementary Figure \ref{FigS14T}(f)] is observed at 4 T, compared to 5 T in the earlier sample [Figure 3(b)]. This upturn occurs 1 T above the magnetic saturation field, effectively ruling out demagnetization effects, which are estimated to be approximately 0.1 T based on the data in Figure 3(b). Additionally, a downturn in $\sigma_{zy}$ appears at 13 T (11 T) at 1.8 K (4 K) as seen in Supplementary Figure \ref{FigS14T}(f) [\ref{FigS14T}(g)]. These high field $\sigma_{zy}$ features do not have accompanying features in magnetization, further supporting the conclusions presented in the manuscript.

\begin{figure*}[ht!]
\begin{center}
\includegraphics[width=.7\linewidth]{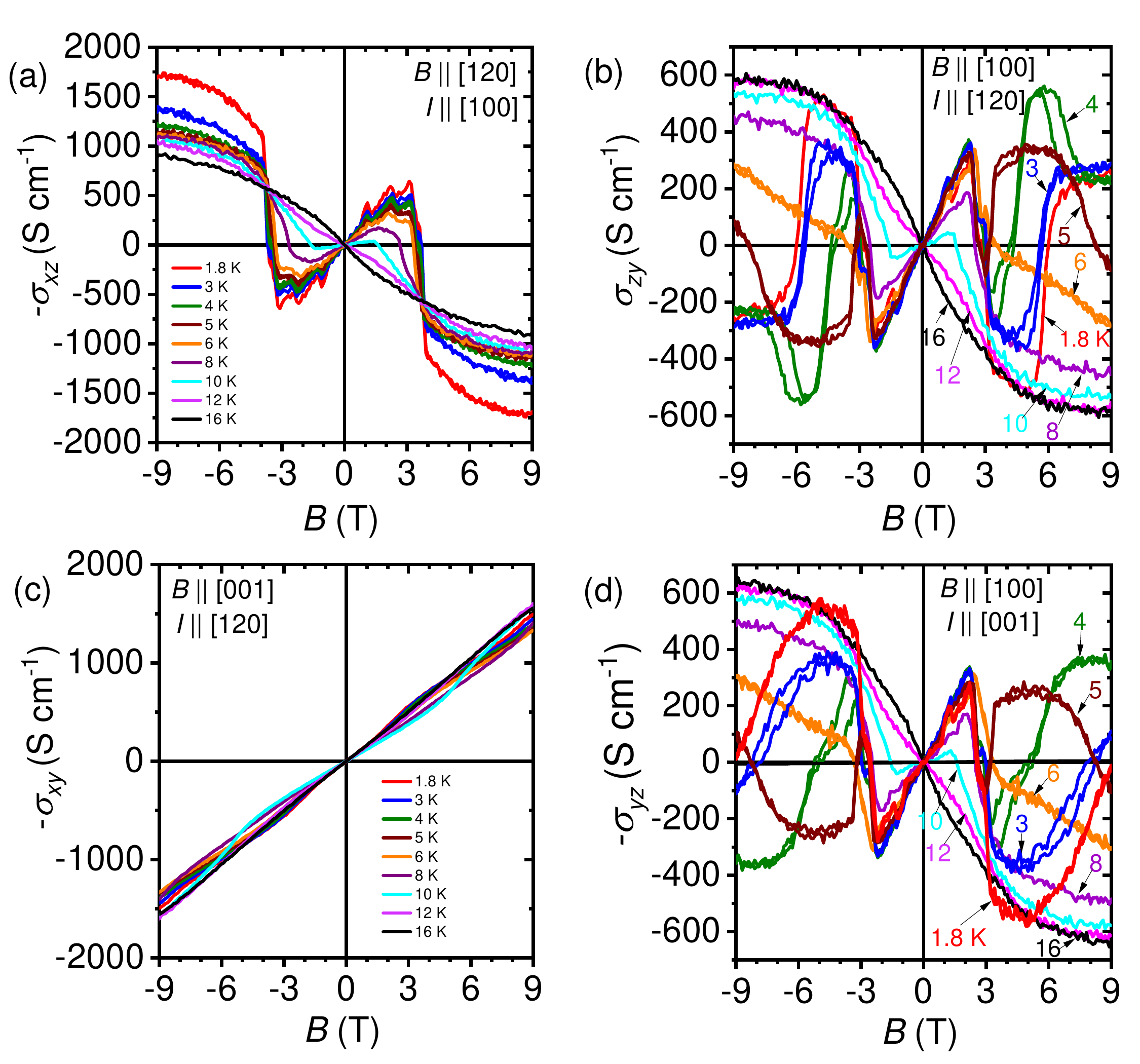}
    \caption{\small \textbf{Hall conductivity vs. magnetic field.} Hall Conductivity of HoAgGe at selected temperatures is plotted as a function of the magnetic field, measured for: (a) $B||[120]$ and $I||[100]$ ($-\sigma_{xz}$), (b) $B||[100]$ and $I||[120]$ ($\sigma_{zy}$,) (c) $B || [001]$ and $I||[120]$ ($-\sigma_{xy}$), and (d) $B||[100]$ and $I||[001]$ ($-\sigma_{yz}$).}
    \label{FigS5}
    \end{center}
\end{figure*}
\newpage

\begin{figure*}[ht!]
\begin{center}
\includegraphics[width=.7\linewidth]{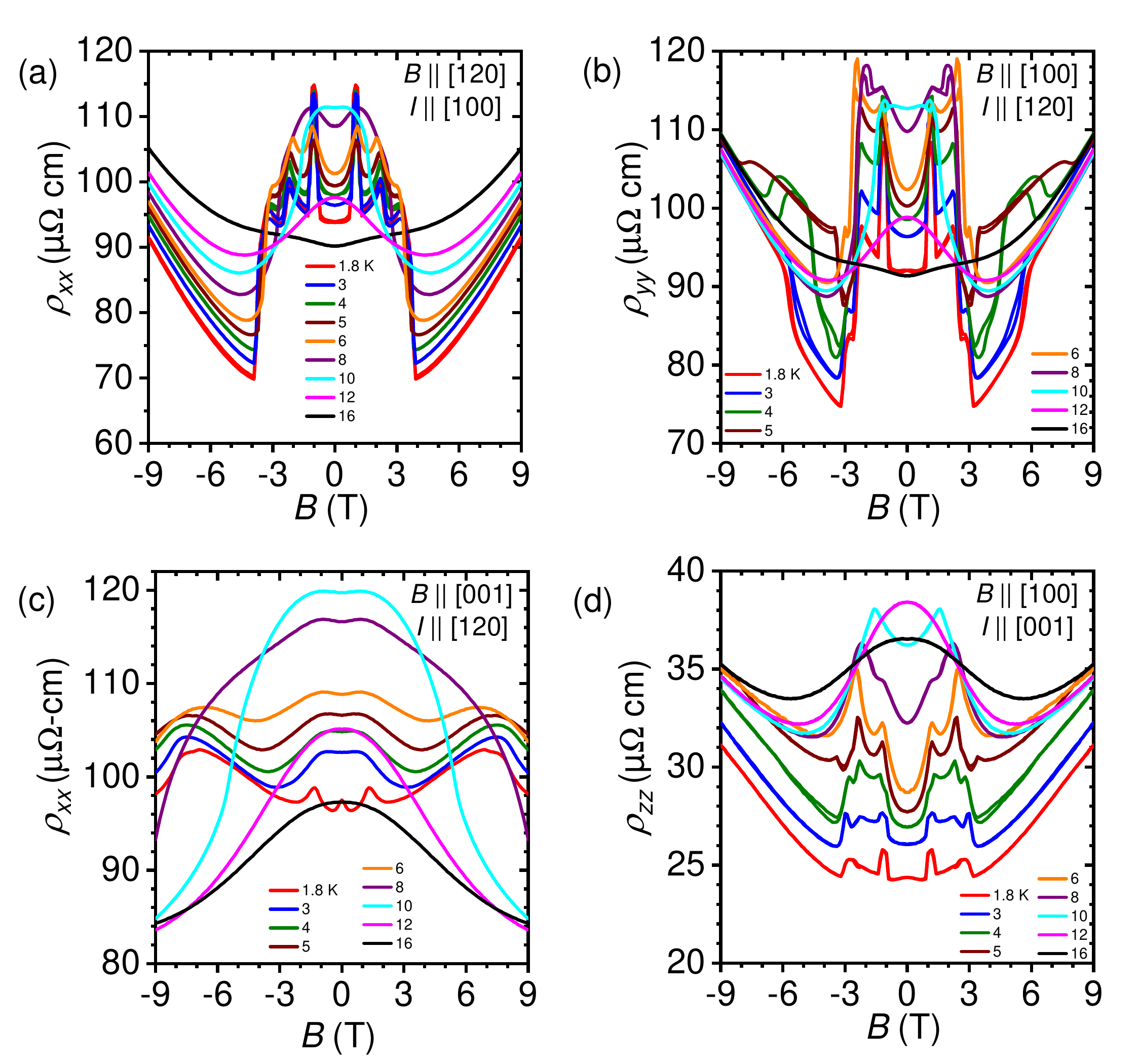}
    \caption{\small\textbf{Resistivity vs. magnetic field.} Longitudinal resistivity of HoAgGe at selected temperatures is plotted as a function of magnetic field, measured for (a) $B||[120]$ and $I || [100]$ ($\rho_{xx}$), (b) $B|| [100]$ and $I || [120]$ ($\rho_{yy}$), (c) $B || [001]$ and $I || [120]$ ($\rho_{yy}$), and (d) $B || [100]$ and $I|| [001]$ ($\rho_{zz}$).}
    \label{FigS6}
    \end{center}
\end{figure*}

\newpage

\begin{figure*}[ht!]
\begin{center}
\includegraphics[width=.7\linewidth]{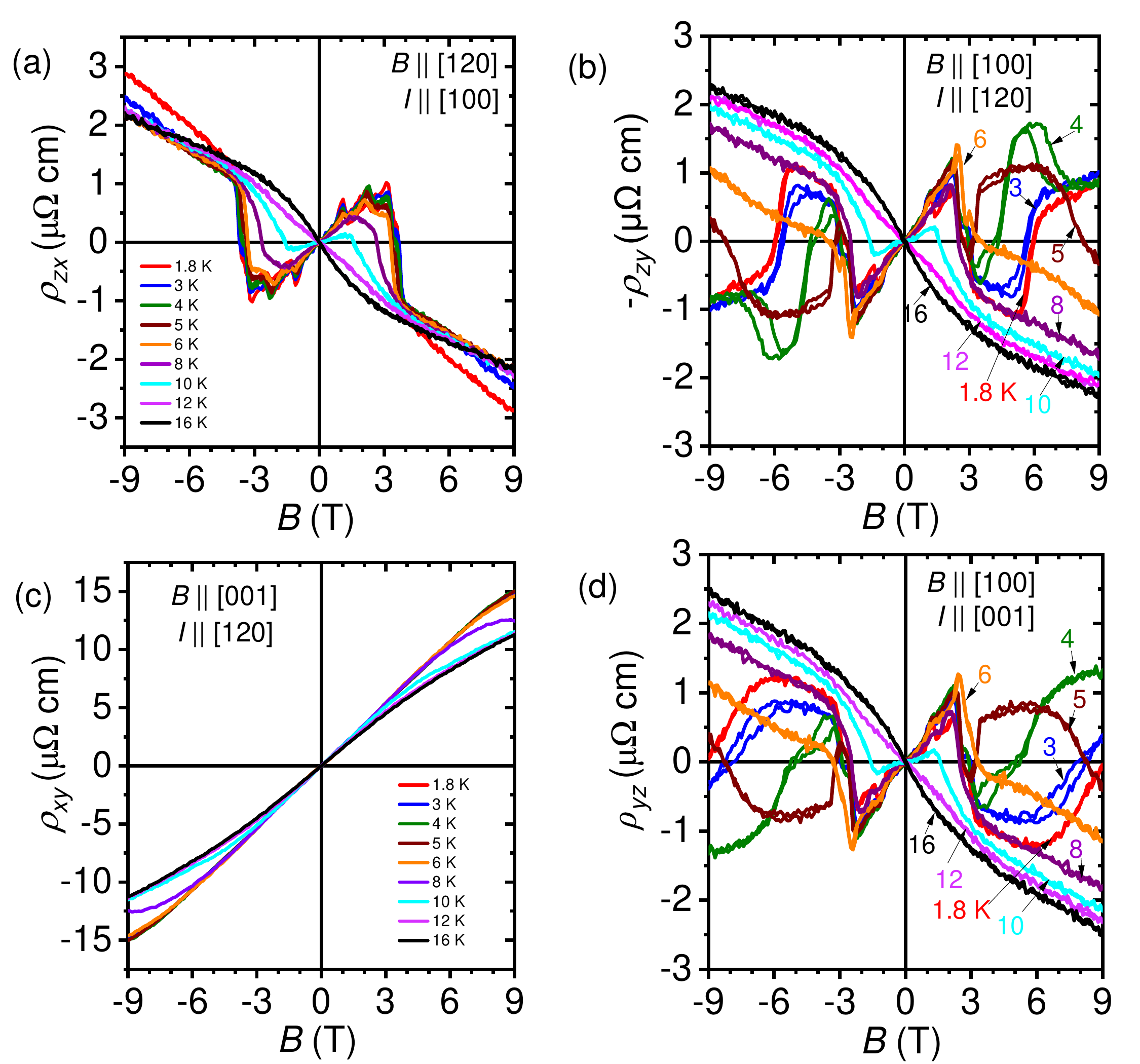}
    \caption{\small \textbf{Hall resistivity vs. magnetic field.} The Hall resistivity of HoAgGe at selected temperatures is plotted as a function of magnetic field, measured for (a) $B||[120]$ and $I || [100]$ ($\rho_{zx}$),  (b) $B|| [100]$ and $I || [120]$ ($-\rho_{zy}$),  (c) $B || [001]$ and $I || [120]$ ($\rho_{xy}$), and (d) $B || [100]$ and $I|| [001]$ ($\rho_{yz}$).}
    \label{FigS7}
    \end{center}
\end{figure*}

\newpage
\begin{figure*}[ht!]
\begin{center}
\includegraphics[width=.6\linewidth]{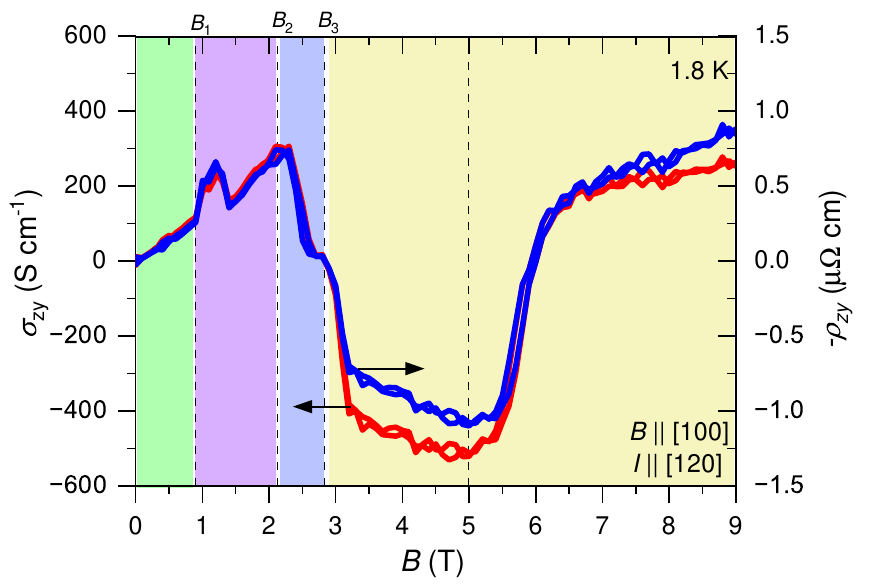}
    \caption{\small \textbf{Hall resistivity and conductivity for magnetic field along [100].} The Hall resistivity ($-\rho_{zy}$) of HoAgGe is plotted on the right axis,  and the Hall conductivity ($\sigma_{zy}$), calculated as $\sigma_{zy}$ = $-\rho_{zy}/(\rho_{zz}\rho_{yy}$), is plotted on the left axis. The plot demonstrates that the steps or jumps in $\rho_{zx}$ and $\sigma_{zy}$ occur at the exact same field. This behavior is unaffected by the fact that $\rho_{zz}$ and $\rho_{yy}$, used in the calculation of $\sigma_{zy}$, were measured on two different samples. Dashed lines and shaded regions in different colors are included as visual guides.}
    \label{FigS8}
    \end{center}
\end{figure*}

\newpage
\clearpage
\begin{figure*}[ht!]
\begin{center}
\includegraphics[width=.37\linewidth]{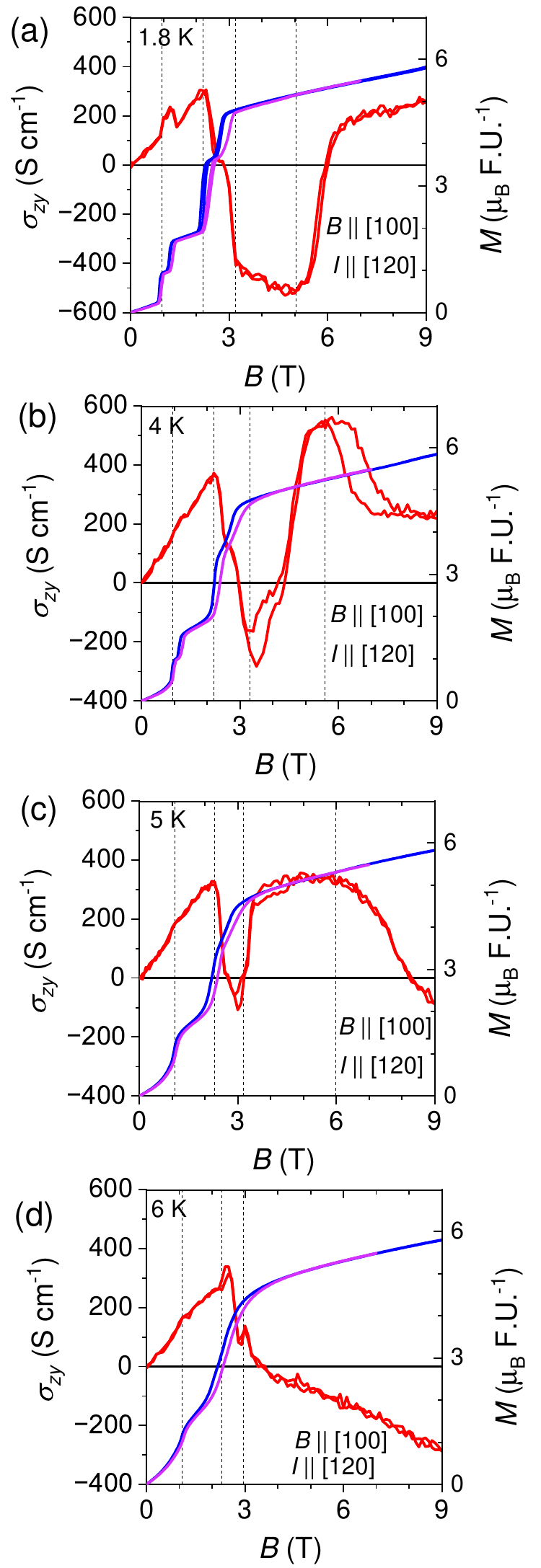}
    \caption{\small \textbf{Hall  conductivity and magnetization for magnetic field along [100].} The Hall conductivity ($\sigma_{zy}$) is plotted on the left axis (red), while the magnetization is plotted on the right axis (blue and pink) for $B||[100]$ at (a) 1.8 K, (b) 4 K, (c) 5 K, and (d) 6 K. The magnetization shown in blue was measured on a relatively larger sample compared to the Hall bar sample. The magnetization shown in pink was measured on the Hall bar sample after detaching the Hall and resistivity contacts and removing the epoxy used to attach them. Dashed lines are included as visual guides.}
    \label{FigS9}
    \end{center}
\end{figure*}

\newpage
\clearpage
\begin{figure*}[!ht]
\begin{center}
\includegraphics[width=.6\linewidth]{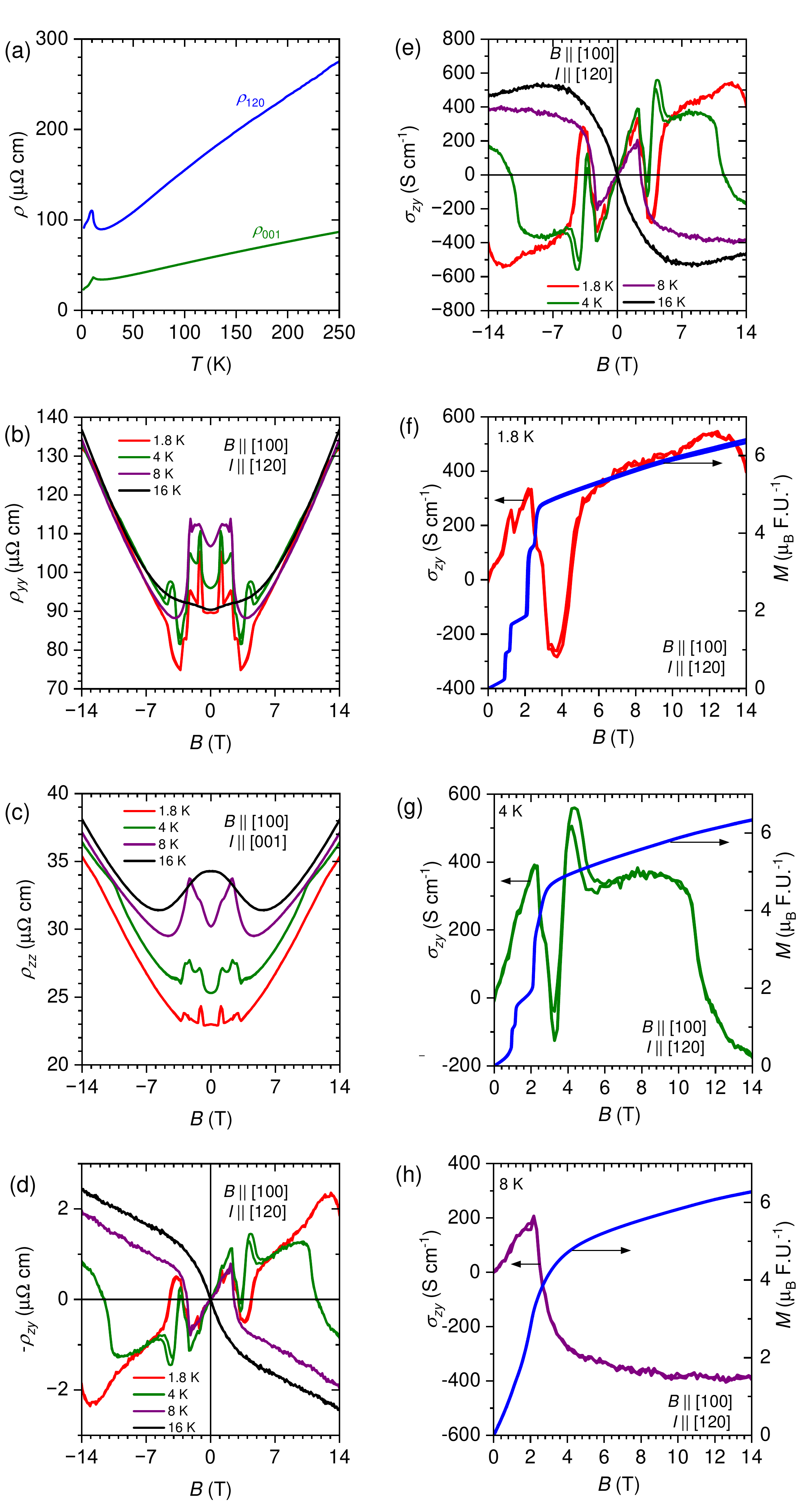}
    \caption{\small \textbf{Magnetotransport of HoAgGe measured up to 14 T on new sets of samples.} a) Resistivity as a function of temperature for $I|| [120]$(blue) and [001](green). b-c) Resistivity as a function of $B||[100]$ at selected temperatures, measured with $I|| [120]$ (b), and $I||[001]$ (c). (d) Hall resistivity ($-\rho_{zy}$) measured with $B|| [100]$ and $I|| [120]$ at selected temperatures. (e) Hall conductivity ($\sigma_{zy}$) for $B|| [100]$ and $I|| [120]$ at selected temperatures. (f-h) $\sigma_{zy}$ and magnetization measured for $B|| [100]$ at 1.8 K (f), 4 K (g), and 8 K (h).}
    \label{FigS10}
    \end{center}
\end{figure*}

\newpage
\clearpage

\begin{figure*}[!ht]
\begin{center}
\includegraphics[width=.5\linewidth]{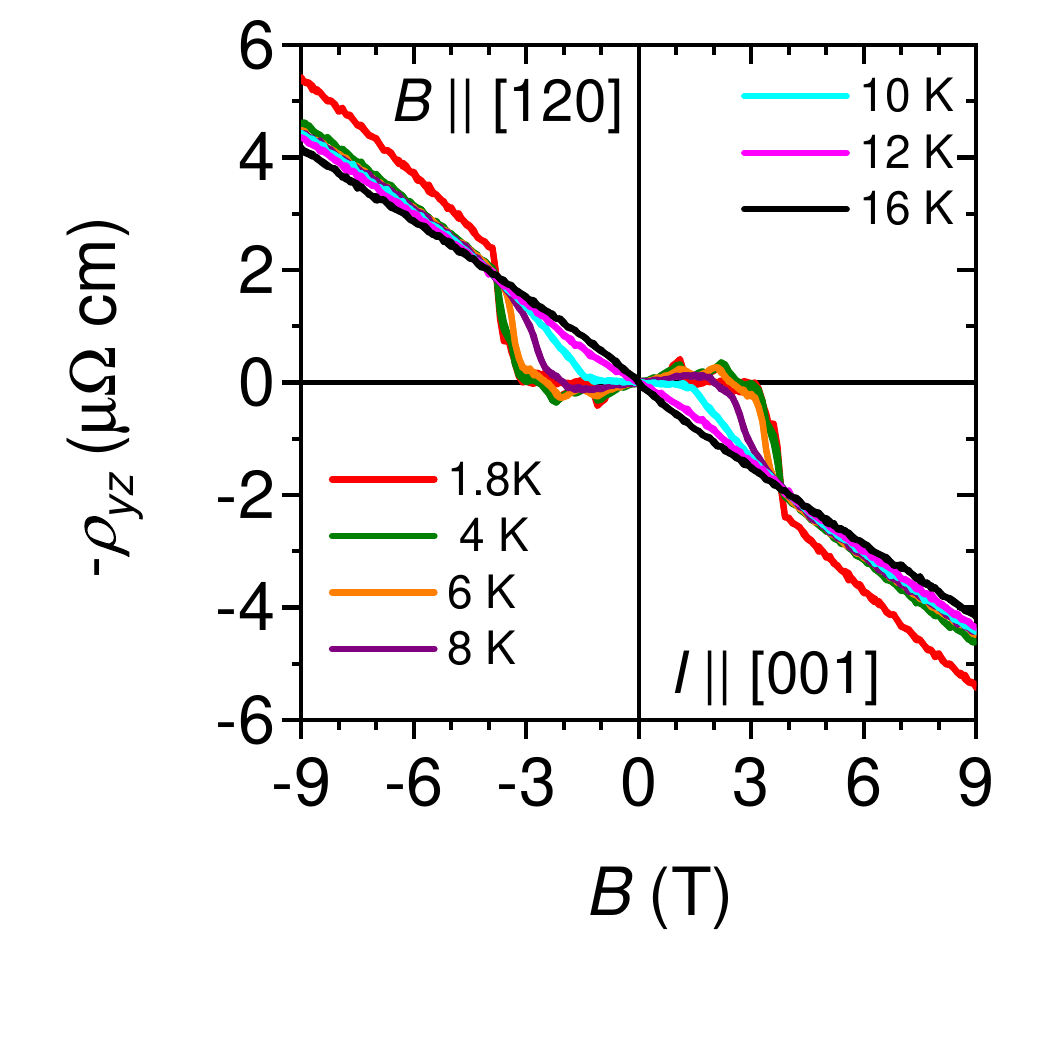}
    \caption{\small \textbf{Hall resistivity for magnetic field along [120] and current along [001].} Hall resistivity as a function of magnetic field applied along [120] at selected temperatures, measured with the current applied along [001].}
    \label{FigS11}
    \end{center}
\end{figure*}

\section{S\lowercase{tructural} c\lowercase{haracterization}}

\begin{figure*}[!ht]
\begin{center}
\includegraphics[width=.6\linewidth]{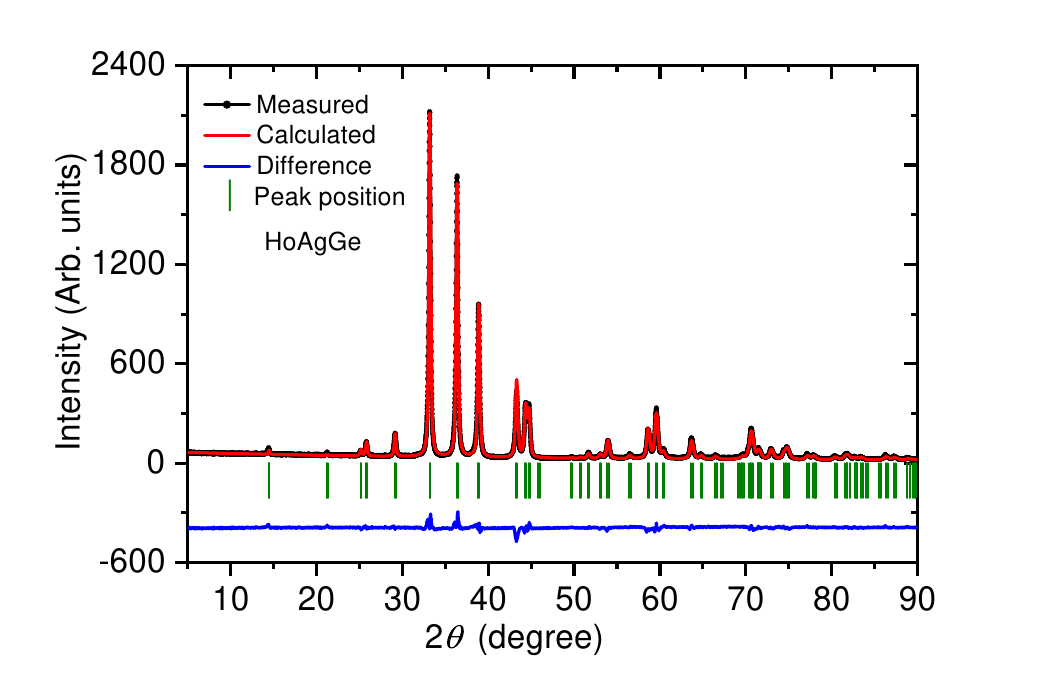}
    \caption{\small \textbf{X-ray diffraction.} Rietveld refinement of the X-ray powder pattern of HoAgGe measured at room temperature.}
    \label{FigS12}
    \end{center}
\end{figure*}

\begin{table}[!ht]
\begin{center}
\caption{Selected data from Rietveld refinement of powder X-ray diffraction collected on ground crystals of HoAgGe. Atomic coordinates are 0.58481, 0, 0 for Ho; 0.24866, 0, $\frac{1}{2}$ for Ag; $\frac{1}{3}$, $\frac{2}{3}$, $\frac{1}{2}$ for Ge(1); and 0, 0, 0 for Ge(2).}\label{T0}
\par%
\begin{tabular}
[c]{ll}\hline
   Space group                           &      \textit{$P\bar{6}2m$} (No. 189)             \\
   Unit cell parameters                &     \textit{a} = 7.0761(7) \AA               \\
                                                    &     \textit{c} =  4.1788(2)  \AA          \\
 \textit{R}$_{WP}$                     &     11.7 \%                               \\
  \textit{R}$_{B}$                     &     5.499 \%                               \\
   \textit{R}$_{F}$                           &     6.778 \%                               \\
 \hline
\end{tabular}
\end{center}

\end{table}


%

\end{document}